# Ultrafast disinfection of SARS-CoV-2 viruses

Yang Xu[#a,b], Alex Wing Hong Chin[#c,d], Haosong Zhong[#a], Connie Kong Wai Lee[#a], Yi Chen[a,b], Timothy Yee Him Chan[a,b], Zhiyong Fan[b], Molong Duan[e], Leo Lit Man Poon[*c,d], Mitch Guijun Li[*a,b]

[a]Division of Integrative Systems and Design, The Hong Kong University of Science and Technology, Clear Water Bay, Kowloon, Hong Kong SAR, P. R. China. Email: mitchli@ust.hk

[b]Department of Electronic and Computer Engineering, The Hong Kong University of Science and Technology, Clear Water Bay, Kowloon, Hong Kong SAR, P. R. China.

[c]School of Public Health, LKS Faculty of Medicine, The University of Hong Kong, 21 Sassoon Road, Hong Kong SAR, P. R. China. Email: llmpoon@hku.hk

[d]Centre for Immunology & Infection, Hong Kong Science Park, Hong Kong SAR, P. R. China

[e]Department of Mechanical and Aerospace Engineering, The Hong Kong University of Science and Technology, Clear Water Bay, Kowloon, Hong Kong SAR, P. R. China.

## Abstract

The wide use of surgical masks has been proven effective for mitigating the spread of respiration diseases, such as COVID-19, alongside social distance control, vaccines, and other efforts. With the newly reported variants, such as Delta and Omicron, a higher transmission rate had been found compared to the ancestral strains. Recent data further suggest that Omicron variant can replicate better than the ancestral virus in the upper respiratoary tract.More frequent sterilization of surgical masks is needed to protect the wearers. However, it is challenging to sterilize the commodity surgical masks with a fast and effective method. Herein, we reported the sterilization of the SARS-CoV-2 viruses within an ultra-short time, while retaining the mask performance. Silver thin film is coated on commercial polyimide film by physical vapor deposition and patterned by laser scribing to form a Joule heating electrode. Another layer of the gold thin film was coated onto the opposite side of the device to promote the uniformity of the Joule heating through nano-heat transfer regulation. As a result, the surgical mask can be heated to inactivation temperature within a short time and with high uniformity. By Joule-heating the surgical mask with the temperature at 90 °C for 3 minutes, the inactivation of the SARS-CoV-2 showed an efficacy of 99.89%. Normal commodity surgical masks can be sterilized faster, more frequently, and efficiently against SARS-CoV-2 viruses and the new invariants.

## Introduction

Wearing surgical masks can reduce the chances of expelling respiratory droplets, which is important to mitigate the spread of respiration diseases during a pandemic.[1, 2] In addition, the pathogens for the respiration diseases are mostly trapped within the respiration droplet, due to the high surface tension of water. The electrostatic potential within the surgical masks can catch the suspended respiration droplets, although the micropores are usually larger than the diameters of the respiration droplets.[3] So, wearing a surgical mask in a highly-risky area can also reduce the chances of breathing in infectious viruses.

With the emergence of the highly infectious new variant of the SARS-CoV-2, it is important to reduce the chance of inhaling infectious aerosols or droplets.[4, 5] A much frequent sterilization has to be performed to inactivate the potential viruses trapped in the surgical mask.[6] Many methods have been reported to sterilize the masks, thereby allowing to extend the lifespan or reuse of this personal protective equipment. UV light is an important process for mask sterilization. The high-energy photons from the UV light can break down the RNA of the virus for a short time. However, the highly overlapping microfibers of the surgical mask block the light from transmitting into the underlying layers. In addition, UV light can also degrade microfibers. The microfibers might break down into small pieces due to the aging of the polymer under the UV.

Heating is another method for sterilizing infectious virus particles. Many functional masks with photothermal capabilities have been developed since the appearance of COVID-19.[7, 8] We have found that the SARS-CoV-2 can be inactivated with 5 min of heating at a 70 °C temperature.[9] Although the oven has high performance and uniformity toward mask heating, the long duration might hinder the wide application of mask sterilization to the public. The commonly used steam sterilization will bring humidity to the mask and decrease the ability of the capturing performances from the electrostatic forces in the microfibers, in addition to its long sterilization time and poor portability.

Herein, we demonstrate an ultrafast portable flexible Joule-heating sterilizer for masks fabricated by laser scribing and made of commercially available polyimide film. The prototype device can sterilize SARS-CoV-2 at a high temperature within a very short time while keeping the mask dry and suitable for wearing.

**Results and Discussion**

Temperature is crucial for the sterilization of the coronavirus. To inactivate all the viruses on the surgical masks, the minimum temperature has to be achieved at every place on the mask. Considering using a planar heating device attached to the surgical masks, the Joule heating will induce non-uniform heating, because the electrical current tends to go through the shortest path to achieve the smallest resistance. So the central part of the heater will achieve a much high temperature as a hot zone, while the edge part has a low temperature due to the Joule heating as a cold zone. Although the heat transfer from the hot zone to the cold zone can decrease the temperature difference, this time-consuming process will slow down the inactivation process and increase the overall processing duration.

To overcome this challenge, we designed a novel geometry to pattern the Joule heating electrode into micro-Joule heating elements. Although the hot zone and cold zone might still exist on the surfaces of the planar heater within each element, the distance between the hot zone and cold zone is significantly shorted. In addition, the Joule heating electrodes are placed at the backside of the substrate. As a result, the generated heat will first go through the polyimide and then onto the surface of the surgical masks. A polyimide substrate is chosen due to its high thermal durability, flexibility, and thermal conductance. As a result, the heat variance between the hot zone and the cold zone is flattened by passing through the polyimide film. From numerical simulations, the variance between the hot zone and the cold zone is reduced to 1 degree.

To further reduce the variance of temperature on the mask, another heat dissipation layer is added. Gold is chosen due to its high biosafety and high thermal conductivity.[10] Its cost can also be reduced by controlling its thickness down to 10 nm. Since the wetting of gold on the polyimide is not good, another 10 nm thick Ti is pre-sputtered on the polyimide film before the deposition of gold. For scalable manufacturing, the low-cost copper can be used as an alternative material to gold. By adding the backside heat dissipation layer, the temperature difference continues to drop to 1 °C. Thus, the overall temperature fluctuation on a 1 cm by 1 cm area is well controlled, as shown in Figure 1.

The Joule heating element is chosen as silver, due to its ultra-high electrical conductivity and thermal conductivity, relatively low cost (compared to gold), and self-sterilization property.[11,12] The laser scribing is used to pattern the microelectrode, due to its high scalability and comability with industrial mass productions.

Traditional heating sterilization options like the high-temperature steam pot consume much time heating the water before the practical sterilization process happens. The vaporization of water also absorbs a large amount of energy during the phase transition, while such energy does not increase the temperature, so the sterilization effect is not improved. This Joule heating device, however, is heated immediately after the switch is turned on, and the heat is transferred from the silver heating element directly to the mask without the phase transition, thus not only saving the time of heating water as the traditional steam sterilizer does but also the extra energy.

When the device is powered by a constant voltage, the heating time until the designed 90 °C is reached still takes around 5 minutes, corresponding to a relatively slow heating speed, mainly because of the limited heating power. However, if a larger power is used, the stable temperature will increase accordingly, and cause the microfibers in the mask to break down. To rapidly heat the device while keeping a reasonably stable temperature, the active heating power is significant, which is realized by the PID control of the applied voltage, as shown in Figure 2. A tiny thermistor is integrated on the surface of the heating element, serving as the temperature sensor that gives negative feedback to the controlling program written on PCB so that the voltage can be adjusted based on the measured temperature. The PID program will first output a relatively high voltage to keep the temperature increasing as soon as possible, and then output a relatively low voltage to slow down the heating or even cool down if the temperature is higher than the target. This adjustment process is repeated on a smaller scale and finally stabilizes the temperature to the designed one.

In addition to the power supply, the testing sample is kept in a 3D-printed container with polystyrene (PS) pieces as the adiabatic layers, which is a common case in heat transfer experiments. The PS adiabatic layer helps maintain the heat generated by the silver heating element because of its poor heat conductivity, and it also prevents the thermal convection with the air from dissipating heat into the surrounding environment.[13] This adiabatic encapsulation increases the stable temperature from 85 °C to 90 °C.

As shown in Figure 3, the Joule-heating mask sterilizer can efficiently inactivate infectious SARS-CoV-2. We make an integrated mask sterilization device for the virus test with PID temperature control to realize precise and real-time temperature monitoring and stabilization. The SARS-CoV-2 (Omicron variant) virus is tested with the device for 3, 5, and 10 minutes. The results clearly show that sterilization at 3 minutes is sufficient to effectively inactivate the SARS-CoV-2 viruses. We propose to further improve the sterilization efficiency by adding another heating layer on the other side of the sample, thus reducing the time of reaching the stabilized temperature.

**Conclusions**

We design a uniform and ultrafast heater and test its sterilization performance against SASRS-CoV-2 viruses. For an Omicron variant, the calibrated heating at 90 oC at 3 min can efficiently inactivate the viruses with 99.89% efficiency. This will lay the foundation for building next-generation portable sterilization devices against COVID-19 and other infectious diseases using heating.

**Figures**

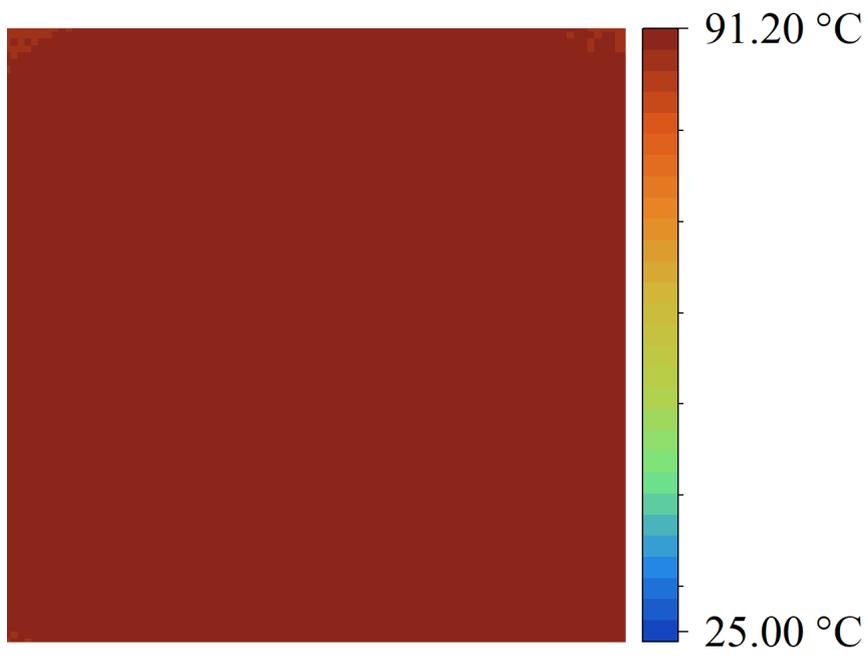

Figure 1: Uniformity of the heating element.

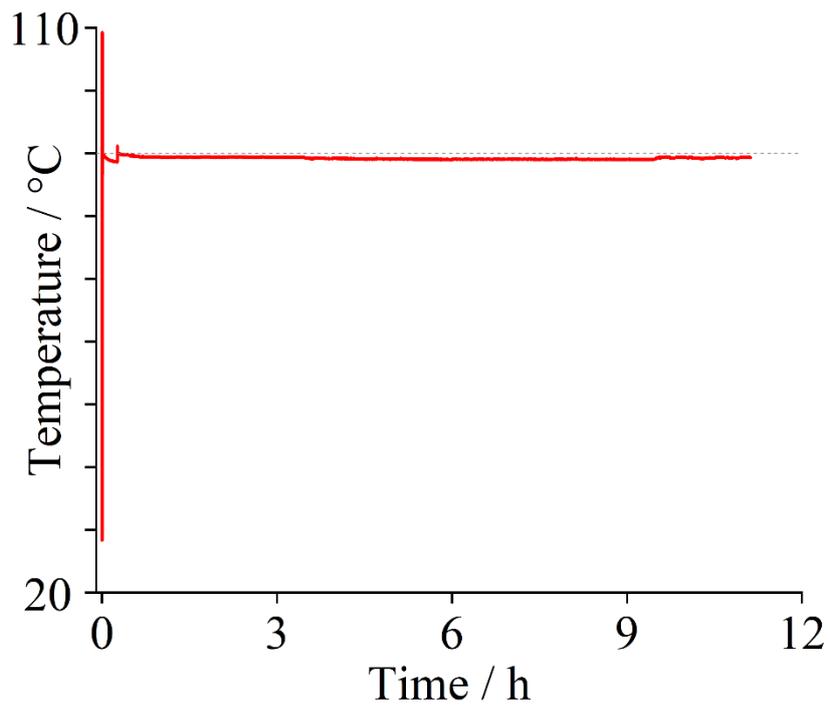

Figure 2: Temperatuer and time relation controlled by the PID

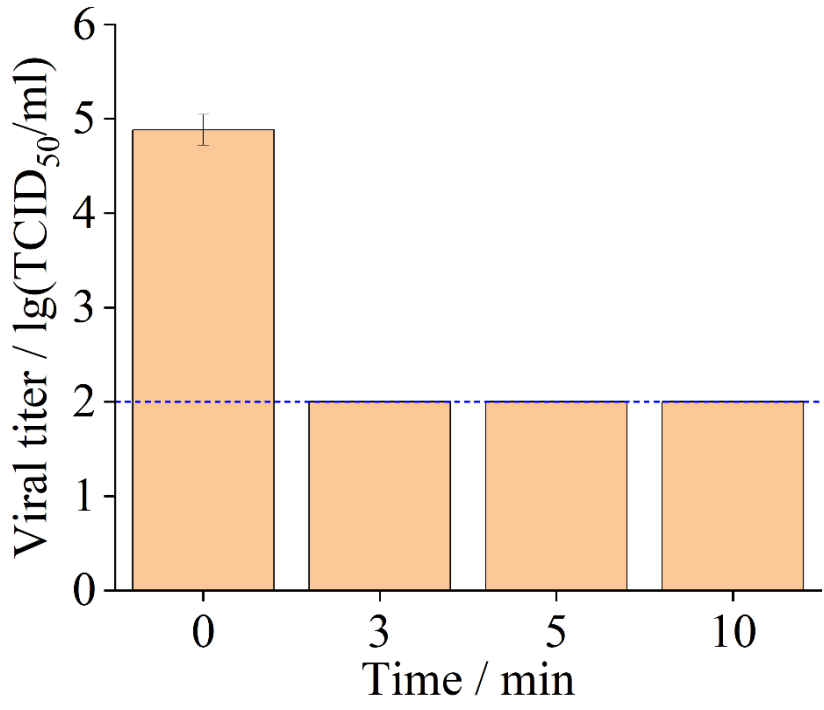

Figure 3: Virus test resolution indicating the inactivation performance.

**Experimental**

The 50 μm thick polyimide film is purchased from Chenxi electrical materials. The polyimide film was used directly after the cleaning process with acetone, methanol, and DI water. One side of the film was deposited with 20 nm of Ti adhesion layer and 70 nm of a silver layer with a Denton magnetic sputter. After laser scribing, we use alcohol and DI water to clean the surface of the scribed sample. After cleaning, the other side of the polyimide is coated with 10 nm of Au as the heat dissipation layer.

The Joule heating electrode was patterned using a 300 ns laser scribing at 1064 nm wavelength. We shine the laser from the PI side and use the following parameters: power 0.01%, speed 350 mm/s, and filling line spacing 20 μm. The conducting wire width is 950 μm, and the gap width is 50 μm.

A PID controller is used for heating and maintaining the temperature of the viruses loaded mask at 90 °C. The temperature distributions of the mask were characterized using a Fluke IR camera with a fixed distance of 15 cm.

For the inactivation test, the virus-containing droplets with volumes ranging from 10 μL were spotted on the surfaces of the surgical mask. Then, the mask heater was used to heat the samples for different durations all at 90 °C, ranging from 3 min to 10 min. Then, the sample will be immersed in a viral transport medium to elute the virus. The virus within the eluted droplet will be assessed using the 50% tissue culture infective dose ($TCID_{50}$) assay of Vero E6 cells. Different dilutions of the eluted viruses will be tested, and the Vero E6 cells will be exposed to every dilution amount. Then the cells will be incubated for 5 days at 37 °C and 5% $CO_2$. Finally, the virus inactivation will be examined for any cytopathic effect. The virus inactivation at each time point will be calculated according to the cytopathic sign ($TCID_{50}$/mL):

$$\log \text{sterilization} = \text{mean} \left[\log_{10}\left(\frac{\text{control sample}}{\text{units}}\right)\right] - \text{mean}\left[\log_{10}\left(\frac{\text{armor structure}}{\text{units}}\right)\right]$$

$$\%\text{sterilization} = [1 - 10^{-\log \text{sterilization}}] \times 100$$

The numerical simulations were performed using finite element methods using commercial COMSOL packages. We use the built-in electrical field module and the heat-transfer-in-solid module. We import the Joule heating multiphysics filed module accordingly. The boundary condition is set to periodic boundary conditions to simulate a repeating structure. The top and bottom of the model are applied the surface-to-ambient-radiation condition to simulate the heat dissipation from the heated sample to the relatively cold environment with natural convection conditions.

**Acknowledgment**

YX, AC, HZ, and CL contributed equally to this work. This project is funded by the University Grants Committee of Hong Kong under project code NA475, HKUST with fund code F0830, RGC theme-based research schemes (T11-705/21-N) and InnoHK grants for

C2I. We would like to thank the help from the State Key Laboratory of Advanced Displays and Optoelectronics Technologies of HKUST for the thin film deposition.